%% file: article.tex
\documentclass[conference]{IEEEtran}
\input{macros}

\usepackage{cite}
\ifCLASSINFOpdf
\else
\fi
\usepackage{amsmath}
\usepackage{algorithm}
\usepackage{algpseudocode}
\ifCLASSOPTIONcompsoc
 \usepackage[caption=false,font=normalsize,labelfont=sf,textfont=sf]{subfig}
\else
 \usepackage[caption=false,font=footnotesize]{subfig}
\fi
\usepackage{graphicx}
\usepackage{todonotes}
\usepackage{amsfonts}
\usepackage{hyperref}
\hypersetup{
    colorlinks=true,
    linkcolor=blue,
    filecolor=magenta,      
    urlcolor=cyan,
    pdftitle={},
    pdfpagemode=FullScreen,
    }

\usepackage{xcolor}

\usepackage{todonotes}  
\usepackage[final]{changes}

\usepackage{amsmath}
\usepackage{amssymb}
\usepackage{amsfonts}
\usepackage{mathtools}
\usepackage{amsthm}
\usepackage{multicol}

\newtheorem{prop}{Proposition}

\theoremstyle{definition}
\newtheorem{definition}{Definition}

\allowdisplaybreaks

\begin{document}
\title{Optimal Wiener-Filter Solutions for Denoising of Graph Signals on Directed Graphs}

% \author{Chun~Hei~Michael~Chan$^{1,2}$, Alexandre~Cionca$^{1,2}$, Dimitri~Van~De~Ville$^{1,2}$
% \\
% $^{1}$ Neuro-X Institute, Ecole Polytechnique Fédérale de Lausanne (EPFL)\\
%   $^{2}$ Department of Radiology and Medical Informatics, University of Geneva}

\author{
\IEEEauthorblockN{Chun Hei Michael Chan, Alexandre Cionca, Dimitri Van De Ville}
\IEEEauthorblockA{
Neuro-X Institute, École Polytechnique Fédérale de Lausanne (EPFL)\\
Department of Radiology and Medical Informatics, University of Geneva\\
Email: \{chunheimichael.chan, alexandre.cionca, dimitri.vandeville\}@epfl.ch
}
}

{}



\maketitle

\begin{abstract}
Graph signal processing has opened new avenues to the canonical denoising problem in interesting settings. Specifically, here we propose a Wiener-filter solution for graph signals on directed graphs. Under various stationarity assumptions combining uncorrelated and correlated noise conditions, we show optimal solutions, including a successful proof-of-concept for temperature graph.
\end{abstract}

% \begin{IEEEkeywords}
% Graph signal processing, Fourier transform, graph signal denoising, Wiener filtering
% \end{IEEEkeywords}

% \fi
\IEEEpeerreviewmaketitle

\section{Introduction}

\IEEEPARstart{T}{he} processing of signals defined on irregular domains arises in many application areas. By modeling such domains with graphs, the analysis and processing of signals on irregular structures fall within the field of graph signal processing (GSP)~\cite{ortega_graph_2018}. Among the various tasks addressed in GSP, signal denoising is a core component of many processing pipelines~\cite{chen_signal_2014, waheed_graph_2018}. 

In \cite{iraji_wss_2025}, a directed graph Wiener filter is proposed to denoise measurements corrupted by directed graph white noise (WN), which is a structured noise dictated by the underlying graph. However, in many practical scenarios, observations are also affected by sensor measurement errors, often modeled by additive decorrelated white noise \cite{oppenheim_signals_1997}. 

Here, we consider an observation model that combines structured noise with additive measurement noise. We propose a graph Wiener filtering framework that is robust to handling both decorrelated additive noise and directed-graph WN. Interestingly, unlike conventional Wiener filters \cite{oppenheim_signals_1997, perraudin_stationary_2017, iraji_wss_2025}, the resulting spectral filter is complex-valued. Finally, we validate the proposed approach through experiments on a real-world graph.

\section{Graph Fourier Transform}
Starting from a directed graph with $N$ nodes represented by the (non-symmetric) adjacency matrix $\ma A$, we assume $\ma A$ is diagonalizable either because application graphs are typically dense \cite{sevi_harmonic_2023} or because it can be diagonalized with minimal perturbation \cite{seifert_digraph_2021}. Its eigendecomposition is
\begin{equation*}
    \ma A = \ma U \ma \Lambda \ma U^{-1},
\end{equation*}
where the eigenvalues $\ma \Lambda[k,k]$ are real or occur in complex-conjugate pairs with corresponding eigenvectors. For a real graph signal $\vc x$, the graph Fourier transform (GFT) is
\begin{equation*}
    \hat{\vc x} = \text{GFT}\{\vc x\} = \ma U^{-1}\vc x,
\end{equation*}
which also yields conjugate coefficient pairs, i.e., \(\hat{\vc x}[k_1] = \hat{\vc x}[k_2]^\star\). Spectral filtering is then defined as
\begin{equation*}
    \vc y = \ma U \hat{\ma G} \ma U^{-1} \vc x,
\end{equation*}
where \(\hat{\ma G} \in \mathbb{C}^{N \times N}\) is a diagonal matrix.

\begin{figure*}[hbt!]
    \centering
    \subfloat[Signal and graph of interest \label{usa-graph}]{
      \raisebox{64pt}{% <-- tune this value
    \parbox{0.33\linewidth}{%
      \includegraphics[width=1\linewidth]{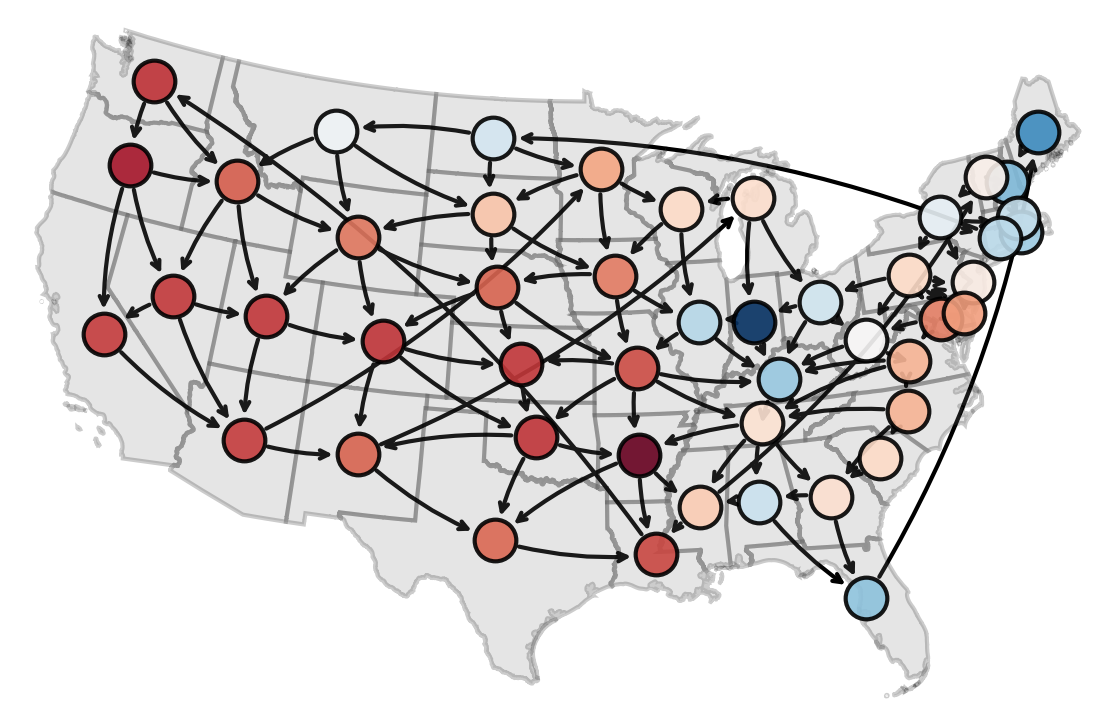}%
      %\leavevmode\vspace*{0pt}
    }%
    }}
  \subfloat[Performance comparison \label{denoising-one-sample}]{%
       \includegraphics[width=0.32\linewidth]{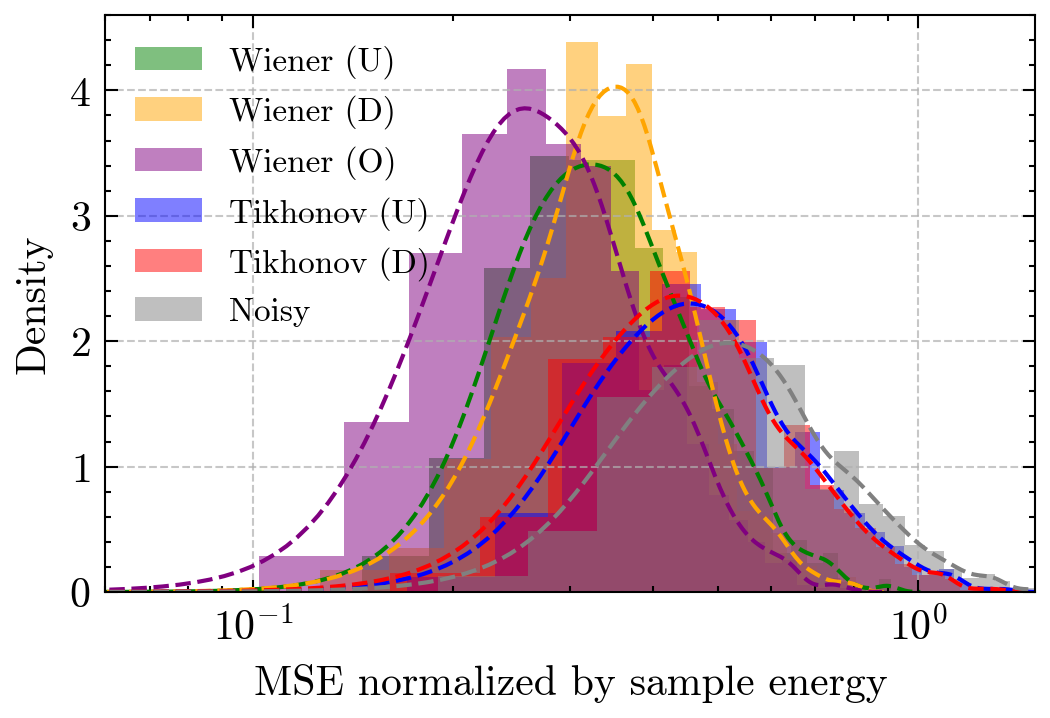}}
    \subfloat[Influence of SNR\label{landscape-denoising-snr}]{%
       \includegraphics[width=0.33\linewidth]{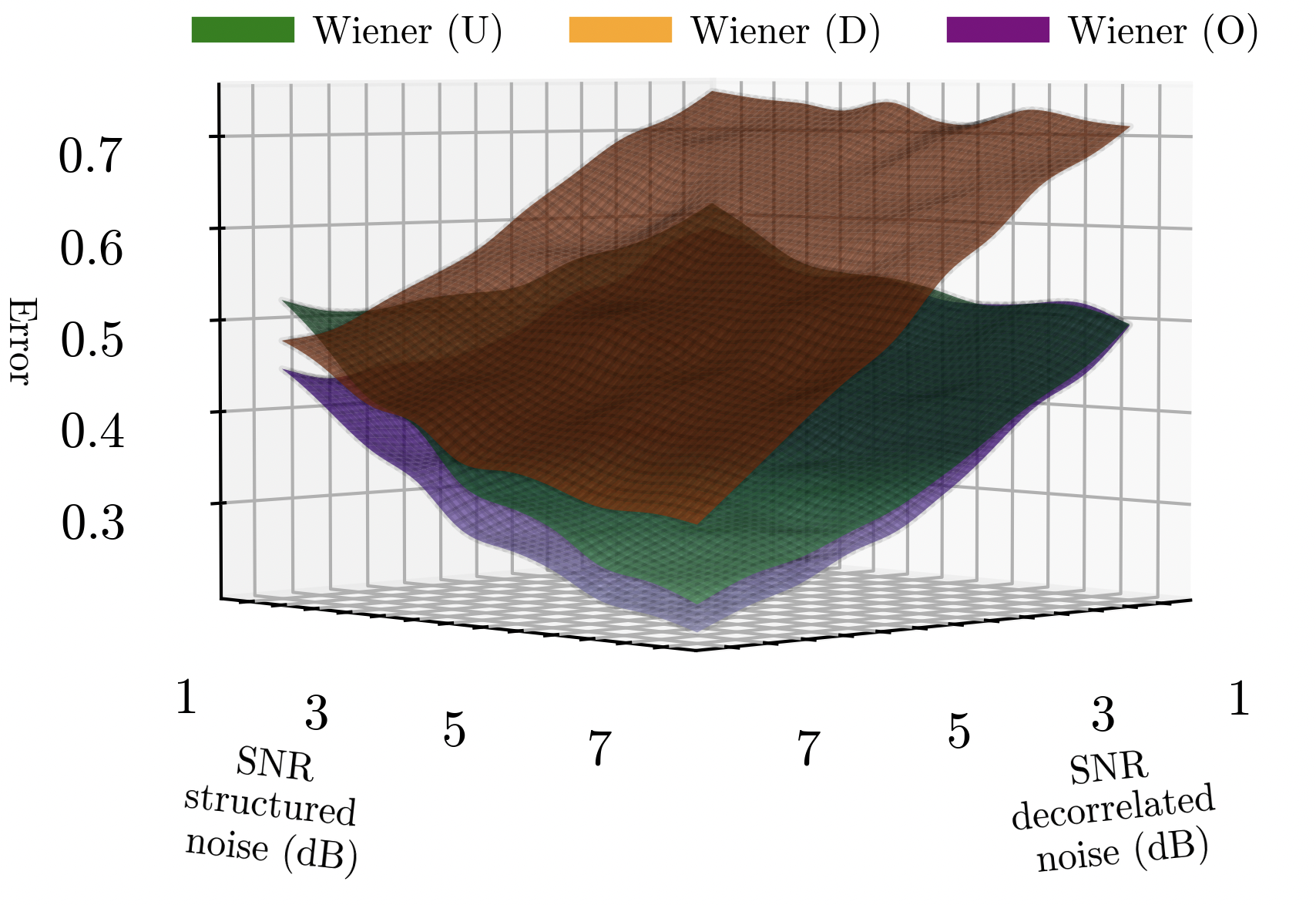}}
  \caption{a) Denoising is carried out on signals residing on the USA graph. b) Denoising performance of undirected (U), directed (D), optimal (O) Wiener, undirected (U), directed (D) Tikhonov ($\lambda =0.01$), average estimator (Noisy) for fixed noise levels $a, b$. c) Denoising results' landscape (RMSE) over varying structured and independent noise levels (here only undirected and directed Wiener are shown as comparison for better visualization of the landscape) .}
  \vspace{0pt}
\end{figure*}

\section{Directed Graph Wiener Filter}

\subsection{Graph Wide Sense Stationary Signal}
A key concept to define the Wiener filter \cite{oppenheim_signals_2017} is stationarity. Therefore we recall the definition of a wide-sense stationary signal on directed graph (DGWSS) \cite{marques_stationary_2017, perraudin_stationary_2017, iraji_wss_2025}. Let $\vc x \in \mathbb{R}^N$ be a random graph signal with mean $\vct \mu_{\vc x}[i]=\mathbb{E}\{\vc x[i]\}$ and covariance $\ma H_{\vc x}=\mathbb{E}\{(\vc x-\vct\mu_{\vc x})(\vc x-\vct\mu_{\vc x})^H\}$. Without loss of generality, we will assume throughout that all random vectors are of mean $\vc 0$.

\begin{definition}
A signal is DGWSS \cite{iraji_wss_2025} if its covariance matrix is can be expressed as,
\begin{align*}
    \ma H_{\vc x} = \ma U \hat{\ma H}_{\vc x} \ma U^H,
\end{align*}
where $\hat{\ma H}_{\vc x}$ is diagonal. Its diagonal entries $\vct\gamma_{\vc x}[k]=\hat{\ma H}_{\vc x}[k,k]$ define the graph power spectral density (PSD). Equivalently, $\hat{\ma H}_{\vc x} = \ma U^H \ma H_{\vc x} \ma U = \mathbb{E}\{\hat{\vc x}\hat{\vc x}^H\}$ being diagonal, shows that the spectral components of $\vc x$ are decorrelated.
\end{definition}

\subsection{Graph White Noise}
Analogous to time-domain WN \cite[Chap. 8.10]{shynk_probability_2013}, a directed graph WN $\vct \epsilon_s$ \cite{chan2026statistical} is DGWSS and is defined through its diagonal graph spectral covariance matrix $\hat{\ma H}_{\vct \epsilon_s}=\ma I_N$, therefore with the following covariance matrix: 
\begin{align*}
    \ma H_{\vct \epsilon_s} = \ma U\hat{\ma H}_{\vct \epsilon_s}\ma U^H= \ma U\ma U^H.
\end{align*}
It should be noted that, despite $\ma U$ being a matrix of complex-valued eigenvectors, $\ma H_{\vct \epsilon_s}$ is a proper covariance matrix for $\vct \epsilon_s$, that it is real-valued and positive semi-definite.

\subsection{Graph Wiener Filter}
Under the assumption of stationary additive noise, the Wiener filter offers a mean-square error optimal estimate of the underlying stationary signal. Here we suggest an optimal graph Wiener to simultaneously remove wide-sense stationary (WSS) noise and DGWSS noise that are structured according to the directed graph. The observation model is defined as follows,
\begin{equation} \label{eq:observation_model}
    \vc y = \vc x + \vct \epsilon_s + \vct \epsilon_d, % \tag{observation model}
\end{equation}
where $\vc x, \vct \epsilon_s$ are DGWSS, and $\vct \epsilon_d$ is WSS. Without loss of generality, let $\mathbb{E}\{ \vc x \}=\mathbb{E}\{ \vct \epsilon_s \}=\mathbb{E}\{\vct \epsilon_d\}={\vc 0}$ i.e signal and noises to be of mean $\vc 0$. 

Performing denoising of $\vc y$ leads to the following formulation of a graph Wiener filter.
\begin{prop}
    The graph Wiener spectral filter $\hat{\ma G}_{\text{opt}}$ is determined by 
    \begin{equation} \label{eq:optimal-wiener}
    \text{diag}\left(\hat{\ma G}_{\text{opt}}\right) = \left(\hat{\ma H}_\vc x + \hat{\ma H}_{\vct \epsilon_s} + \ma M\right)^{-1}\vct \gamma_\vc x,
    \end{equation}
where $\ma M = \left((\ma U^{-1}\ma H_{\vct \epsilon_d} \ma U^{-H})\circ (\ma U^H\ma U)^T\right)$, is an optimal linear filter w.r.t. the expected quadratic error between the filtered observations and $\vc x$.
\end{prop}

\begin{prop}
    $\hat{\ma G}_{\text{opt}}$ is complex-valued and all real signal filtered by $\hat{\ma G}_{\text{opt}}$ remain real.
\end{prop}

While Wiener filters are generally real-valued \cite{oppenheim_signals_2017, perraudin_stationary_2017, iraji_wss_2025}, the proposed filter is complex-valued. Consequently, the denoising operation jointly performs amplitude modulation in the spectral domain and a phase shift \cite{chan2025hilbert}. Intuitively, the phase shift is required to ``synchronize'' WSS noise and DGWSS noise to then perform dampening. Moreover, for a real-valued graph signals, the spectral filter $\hat{\ma G}_{\text{opt}}$ preserves conjugate relationship of spectral coefficient pairs, thereby ensuring that the filtered graph signal remains real-valued \cite{chan2025hilbert}. 

% In the case where WSS noise is absent, $\hat{\ma G}_{\text{opt}}$ reverts to the directed graph Wiener. 

\section{Experiments}
Starting from the observation model in~\eqref{eq:observation_model}, we define the random graph signals
\begin{gather*}
\vc x\sim \mathcal{N}(\vc 0, a^2\ma U\hat{\ma H}_\vc x \ma U^{H}),\\
\vct \epsilon_s\sim \mathcal{N}(\vc 0, b^2\ma U\ma U^{H}), \, \text{(structured WN)},\\
\vct \epsilon_d\sim \mathcal{N}(\vc 0, c^2\ma I_N), \, \text{(decorrelated WN)},
\end{gather*}
with $\vc x, \vct \epsilon_s$ that are DGWSS w.r.t the USA graph (Fig.~\ref{usa-graph}) \cite{seifert_digraph_2021}, $\vct \epsilon_d$ is WSS, and all scaled by respectively $a,b,c \in \mathbb{R}$. Notably, $\vct \epsilon_s$ and $\vct \epsilon_d$ are respectively directed graph WN and decorrelated WN. As for the signal of interest we choose the spectral covariance of $\vc x$ to be such that $\vct \gamma_\vc x[k]=e^{-k}$. We show an example of a smooth $\vc x$ in Fig.~\ref{usa-graph}. 

In the following experiments, we compare the denoising performance of our optimal Wiener filter from~\eqref{eq:optimal-wiener} with that of undirected (i.e., symmetrization of the graph) \cite{perraudin_stationary_2017}, directed graph Wiener \cite{iraji_wss_2025}. We also compare against the output of the undirected, directed Tikhonov \cite{perraudin_stationary_2017, iraji_wss_2025}, and the mean of the observations across noise samples as an estimate for each sample of $\vc x$.

We show in a first instance the denoising results for specific parameters set to $a=22, b=2.5, c=1$ where we sample $1000$ times from $\vc x$ and for each signal sample we generate 100 observations by adding noise samples of $\vct\epsilon_s$ and $\vct\epsilon_d$. The relative mean squared error (RMSE) is shown for each sample in Fig.~\ref{denoising-one-sample}. We observe an overall lower RMSE in the optimal Wiener relatively to undirected and directed counterparts. Moreover, the existence of i.i.d noise largely perturbs performance of directed graph Wiener to the extent that the undirected graph Wiener outperform directed graph Wiener despite a DGWSS signal of interest. 

In a second instance, we keep signal strength to $a=22$ and vary $b, c$ (Fig.\ref{landscape-denoising-snr}) such that SNR ranges between $[1, 7]$ dB. We observe that our optimal graph Wiener outperforms all other denoisers and especially both undirected and directed graph Wiener for all levels of noise. Additionally, we notice that increasing decorrelated noise strongly perturbs performance of directed graph Wiener, once again confirming the need for our optimal Wiener. For both experiments, the implementations are provided in this GitHub repository\footnote{\url{https://github.com/MIPLabCH/DirectedGraphWiener}}.

Finally, we show that acknowledging existence of mixed structured and decorrelated WN allows a more robust directed graph Wiener filter definition. Specifically, we showed that directed graph Wiener filter \cite{iraji_wss_2025} is sensitive to decorrelated noise, naturally present in real data in the form of measurement noise which is overcome by our proposed optimal graph Wiener. 

\pagebreak

\ifCLASSOPTIONcaptionsoff
  \newpage
\fi



\bibliographystyle{ieeetr}



%\vfill


\bibliography{references}

\end{document}

%% file: macros.tex
\newcommand{\vc}[1]{{\mathbf #1}}
\newcommand{\vct}[1]{\pmb{#1}}
\newcommand{\ma}[1]{{\mathbf #1}}

\newlength{\subcolumnwidth}

\newcommand{\nextsubcolumn}[1][]{%
  \cr\noalign{\hfill}
  \if\relax\detokenize{#1}\relax\else\hsize=#1\setlength{\subcolumnwidth}{\hsize}\fi
}